\documentclass[twocolumn]{article}
\usepackage[T1]{fontenc}
\usepackage{graphicx}
\usepackage{epsf,rotate}
\usepackage{epsfig}
\usepackage{rotating}
\usepackage{cite}

\begin{document}

\title{Short-Time Critical Dynamics of Damage Spreading in the Two-Dimensional Ising Model}
\author{M. Leticia Rubio Puzzo\footnote{email adress:lrubio@inifta.unlp.edu.ar} and Ezequiel V. Albano\footnote{homepage:www.gfc.inifta.unlp.edu.ar}}
\date{Instituto de Investigaciones Fisicoqu\'{i}micas Te\'oricas y Aplicadas (INIFTA), UNLP,CCT La Plata - CONICET, c.c. 16, Suc. 4, 1900 La Plata, Argentina}
\maketitle

\begin{abstract}
The short-time critical dynamics of propagation of damage in the Ising ferromagnet in two dimensions is studied by means of Monte Carlo simulations. 
Starting with equilibrium configurations at $T= \infty$ and magnetization $M=0$, an initial damage is created by flipping a small amount of spins in one of the two replicas studied. In this way, the initial damage is proportional to the initial magnetization $M_0$ in one of the configurations upon quenching the system at $T_C$, the Onsager critical temperature of the ferromagnetic-paramagnetic transition. 
It is found that, at short times, the damage increases with an exponent $\theta_D=1.915(3)$, which is much larger than the exponent $\theta=0.197$ characteristic of the initial increase of the magnetization $M(t)$. 
Also, an epidemic study was performed. It is found that the average distance from the origin of the epidemic ($\langle R^2(t)\rangle$) grows with an exponent $z^* \approx \eta \approx 1.9$, which is the same, within error bars, as the exponent $\theta_D$.
However, the survival probability of the epidemics reaches a plateau so that $\delta=0$.
On the other hand, by quenching the system to lower temperatures one observes the critical spreading of the damage at $T_{D}\simeq 0.51 T_C$, where all the measured observables exhibit power laws with exponents $\theta_D = 1.026(3)$, $\delta = 0.133(1)$, and $z^*=1.74(3)$.

\end{abstract}



\section{Introduction}
\label{intro}

The critical behavior of a statistical system near a second-order phase transition is characterized by critical exponents defining a universality class \cite{gold}. 
This behavior is due to the fact that the characteristic length of the physical system given by the spatial correlation length ($\xi$) becomes infinite at $T_C$, the critical point, as $(T-T_C)^{-\nu}$, where $\nu$ is the correlation length exponent, leading to the observation of scale invariance.
But at $T=T_C$, the correlation length grows according to $\xi (t) \propto t^{1/z}$, where $z$ is the dynamic critical exponent, so for a finite system of side $L$ and large enough time, $\xi$ saturates at a certain correlation time $\tau$, such that $\tau \propto L^z$.
Because $z>0$, the correlation time increases rapidly with the lattice size, and therefore, in Monte Carlo simulations it is very difficult to generate independent configurations.
This effect, known as critical slowing down, affects not only measurements in equilibrium but also measurements of dynamic observables.
In this context, the development of the short-time dynamic (STD) theory \cite{jansenn89} provides a very useful tool not only to avoid the critical slowing down but also for the measurements of critical exponents.
In fact, 20 years ago, using renormalization group techniques, Jansenn \textit{et al.} \cite{jansenn89}  showed that the scaling behavior is not only valid in equilibrium but also that dynamic scaling holds during the short-time regime of the evolution of a critical system.
It is worth mentioning that recent progress in the understanding of phase transitions and critical phenomena has been boosted by a large number of investigations based on the application of STD. 
Indeed, the STD approach has proved to be an invaluable tool for the study of a large variety of equilibrium systems (e.g., models such as the Ising \cite{santos2000, santos2000b, zheng1, zheng2, zheng3, zheng4, luo2001, chen2002, silva2002, bab2005, wanz2006, pini2007}, the \textit{XY} \cite{luo98, luo1, luo2, luo3, chen2001, luo2002, medve2003, costa2005, bekh2006, nie2006, lei2007}, the Heisenberg \cite{zelli2007}, the Potts \cite{silva2002, ying2000, yin2004},  the Blume-Capel \cite{silva2002a, grandi2004}, etc.).
Furthermore, the generalization of the STD scaling approach to far-from-equilibrium \cite{sarac} and self-organized critical \cite{laneri} systems poses a large theoretical challenge.

In order to apply the STD scaling approach in Monte Carlo simulations of a magnetic system, the sample should be prepared at a very high temperature, with a small magnetization $M_0$ remaining. 
It is then suddenly quenched to the critical temperature $T_C$ and released to the selected dynamic evolution of the model (Glauber, Metropolis, heat bath, etc.).
The dynamic scaling relationship obtained by Jansenn et al. \cite{jansenn89}, in the case of systems of finite size, and for the $k$th moment of the magnetization, reads \cite{zheng98}

\begin{equation}
M^{(k)}(t, \varepsilon, L, M_0)=b^{-k\beta/\nu} M^{(k)}(b^{-z}t, b^{1/\nu}\varepsilon, b^{-1}L,b^{x_0}M_0),
\label{eq:magsh}
\end{equation}
where $\varepsilon =1-T/T_C$ is the reduced critical temperature, and $\beta$ and $\nu$ are the equilibrium critical exponents for the order parameter and the correlation length, respectively; $z$ is the dynamic critical exponent and the new independent exponent $x_0$ is the scaling dimension of the initial magnetization $M_0$.

Numerical simulations support the theoretical prediction for the short-time dynamic scaling \cite{huse, humayun, menyhard, li, schulke, grass, zheng98, zheng99}. Further, the short-time dynamic scaling is found to be very general \cite{oerding1, oerding2, oerding4, oerding3,zheng98,zhang99,zheng99b,bray,jensen,luo1, luo2, luo3, ying2000, menyhard96, mendes,tome1,tome98}.

In the thermodynamic limit ($L\rightarrow \infty$) and choosing the scaling factor $b=t^{1/z}$ (so that the first argument of the scaling function on the right-hand side of Eq. (\ref{eq:magsh}) is set to $1$), the magnetization of the system ($k=1$) can be written as
\begin{equation}
M(t)=M_0 t^{\theta} F(t^{1/\nu z}\varepsilon),
\label{eq:magnsh}
\end{equation}
where the exponent $\theta=(x_0 -\beta/\nu)/z$ has been introduced. In most cases $\theta>0$, i.e., for $\varepsilon=0$ the small initial magnetization increases in the short-time region.

Summing up, after a microscopic time such as $\xi (t_{\mathrm{mic}})$ is of the order of the lattice spacing, one observes an initial increase of the magnetization according to Eq. (\ref{eq:magnsh}) up to a mesoscopic time of the order of $t_m \sim M_0^{-z/x_0}$, while for $t>t_m$ the standard relaxation of the order parameter, namely, $M \sim t^{-\beta/\nu z}$, is observed.

On the other hand, 
the damage spreading (DS) method is a standard technique frequently used to study the propagation of perturbations in lattice systems. 
In past decades it has been applied to very different systems such as the Ising magnet, spin glasses, Potts models, cellular automata, biological models, etc. (for more details of this technique, see the recent review \cite{rubio2008} and references therein). 
The methodology used to study the propagation of a perturbation by means of DS is to start the simulations with two well-equilibrated configurations, $S^{A}(T)$ and $S^{B}(T)$, which differ from one another only in the state of a small number of sites. 
Then, both configurations are allowed to evolve in time, with the same sequence of random numbers (in terms of the simulations this condition means that both configurations experience the same thermal noise).  
Thus, at certain time $t$, the difference between $S^A$ and $S^B$ will only be as a consequence of the small initial perturbation or damage introduced in the system at time $t=0$.

In order to measure the difference between these two configurations, the Hamming distance or damage ($D(t)$) is defined \cite{herr92} as

\begin{equation}
D(t)=\frac{1}{2N}\sum ^{N}_{l}\left| S^{A}_{l}(t,T)-S^{B}_{l}(t,T)\right|,
\label{eq:dam}
\end{equation}
where the summation runs over the total number of sites $N$, and the index $l\,(1\leq l\leq N)$ is the label that identifies the sites of the configurations.
$S^{A}(t,T)$ is an equilibrium configuration of the system at temperature $T$ and time $t$, while $S^{B}(t,T)$ is the perturbed configuration \cite{herr92,herr90}.

If the initial perturbation introduced in $S^B$ is small ($D(t=0)\rightarrow 0$), at least two possible scenarios are expected: (i) $D(t\rightarrow \infty)$ goes to a finite
nonzero value, and the perturbation is relevant to the system, or (ii) the small perturbation vanishes after a certain time and $D(t\rightarrow
\infty)\rightarrow 0$.
This behavior introduces a new continuous and irreversible critical transition between a state where damage heals and a state where the perturbation propagates into the system.
This transition is known as damage transition and, in the case of the Ising model in two dimensions with Glauber dynamics, it occurs at a damage critical temperature $T_D=
0.992(2) T_C$ \cite{gras2}, where $T_C$ is the Onsager critical temperature for the paramagnetic-ferromagnetic transition.

The
universality class of the damage transition is still an open question.
Grassberger \cite{gras3} conjectures that the DS transition may
belong to the directed percolation (DP) universality class 
\cite{pd5,jensen1,voigt} if its critical point does not coincide
with a critical transition of the physical system, e.g., the
critical temperature of the Ising magnet. 
Numerical simulations of different systems such as the Domany-Kinzel cellular automata, the two-dimensional (2D) Ising model with Swendsen-Wang dynamics \cite{hinrich98}, as well as the
deterministic cellular automata with small noise \cite{bagnoli92} provide evidence that the DS
transition is characterized by critical exponents of the DP
universality class.
However, there are also other systems where the DS transition has
a non-DP behavior, such as the case of the Kauffman model
\cite{kauf69, kauf84} or the Ziff-Gulari-Barshad (ZGB) model in 2D \cite{albanoZGB} (for more details, see \cite{rubio2008}).

STD has been applied to DS in the Ising model at the Onsager
temperature with heat bath dynamics in two and three dimensions by Grassberger
\cite{grass}. 
He found that the damage exhibits power-law behavior of the form
\begin{equation}
D(t)\propto t^{\theta_D},
\label{eq:pwlaw}
\end{equation}
where $\theta_D$ is the initial increase exponent for the damage.
The exponent $\theta_D$ was found
to be $\theta_D=0.191(3)$ ($d=2$) and  $\theta_D=0.104(3)$ ($d=3$).
Also, the dynamic dependence of the survival probability of damage $P(t)$, defined as the probability at time $t$ of finding nonzero damage, was analyzed in \cite{grass}. Grassberger found, as expected \cite{gras89}, a power-law decrease with an exponent $\delta$, of the form
\begin{equation}
P(t) \propto t^{-\delta},
\label{probdelta}
\end{equation}
with $\delta \approx 0.9$ ($d=2$) and  $\delta\approx 1.1$ ($d=3$).
For random independent initial configurations (i.e., half of the spins
damaged on average) a decrease in damage is observed with an exponent $\theta_D'=0.43(2)$ ($d=3$). 

Within this context, the aim of this work is to report results obtained upon the study of the STD of damage spreading in the two-dimensional Ising model with Metropolis dynamics.
In contrast to previous studies \cite{rubio2008} where the initial configuration is equilibrated at a given temperature where the DS is subsequently measured, here we equilibrate the system at $T=\infty$ (i.e., we obtain fully uncorrelated initial configurations), and afterwards we quench the system at the temperature of interest for the measurement of the damage.
Therefore, we expect to gain insight into the propagation of the damage in a quite different scenario than in standard studies, and search for a relationship between damage spreading and the short-time dynamic behavior of relevant observables, e.g. the magnetization. 

The paper is organized as follows. In Sec. \ref{sec2}, we briefly describe the Ising Model and give the simulation details, in Sec. \ref{sec3} we report and discuss the results obtained, and finally in Sec. \ref{sec4}, the conclusions are presented. 

\section{Ising Model}
\label{sec2}

As was mentioned in the previous section, we study the STD of damage spreading in the two-dimensional classical Ising model \cite{ising}. For this reason, in the present section we introduce a brief description of this archetypical system largely used to study phase transitions and critical phenomena in magnetic systems. 

In the Ising model, each site $\sigma _{i}$ of the lattice of size $L\times L$ represents
a spin variable, which interacts with its nearest-neighbor spins with a constant of exchange $J$. If $J>0$, and in the absence of an external magnetic field, the Hamiltonian (${\cal H}$) can be written as

\begin{equation}
{\cal{H}}=-J\cdot \sum _{<i,j>}\sigma _{i}\sigma _{j}, 
\label{hamis}
\end{equation}

\noindent where $\sigma _{i}$ is the Ising spin variable that can
assume two different values $\sigma_i = \pm 1$, the indexes $1\leq
i,j \leq N$ are used to label the spins,  and the summation runs over all the nearest-neighbor pairs of spins. 
In the absence of an external magnetic field and at
low temperature, the system is, for $d > 1$, in
the ferromagnetic phase and, on average, the majority of spins are pointing
in the same direction. In contrast, at high temperature the system
maximizes the entropy, thermal fluctuations break the order and
the system is in the paramagnetic phase. This
ferromagnetic-paramagnetic critical transition is a second-order
phase transition and it occurs at a well-defined critical
temperature ($T_C$). In the two-dimensional case, one has exactly
$k T_C/J = 2/ln(1+\sqrt{2})=2.269...$, where $k$ is the Boltzmann
constant.

With the aim of studying the STD of DS, we start the simulation with a configuration $S^A$ with strictly zero initial magnetization. 
This constraint was introduced by starting the simulation with all spins of configuration $S^A$ equal to $\sigma=1$, and then $N/2$ spins (randomly chosen) of the lattice are flipped to $\sigma=-1$. 
In this context, the initial configuration with uncorrelated spins and $M=0$ corresponds to $T=\infty$.
At $t=0$, a replica $S^B$ was created, and the initial damage was introduced by flipping $N_0$ spins randomly chosen. In this way, the initial damage of the configuration $S^B$ is related to the nonzero initial magnetization as $M_0=2D(t=0)=2 N_0/L^2$.

In order to set the time scale, we assume that during a Monte Carlo time step 
(mcs) all spins of the system ($L\times L$ in total) have the chance 
to be flipped once, on average.

In this context, the spreading of damage is studied by assuming Metropolis dynamics and the simulations were performed on the square lattice of size $L \times L$ ($100 \leq L\leq 4096$), with periodic boundary conditions between the borders of the lattice.

\section{Results and discussion}
\label{sec3}

\begin{figure}[!tbp]
\includegraphics[width=8cm,clip=true]{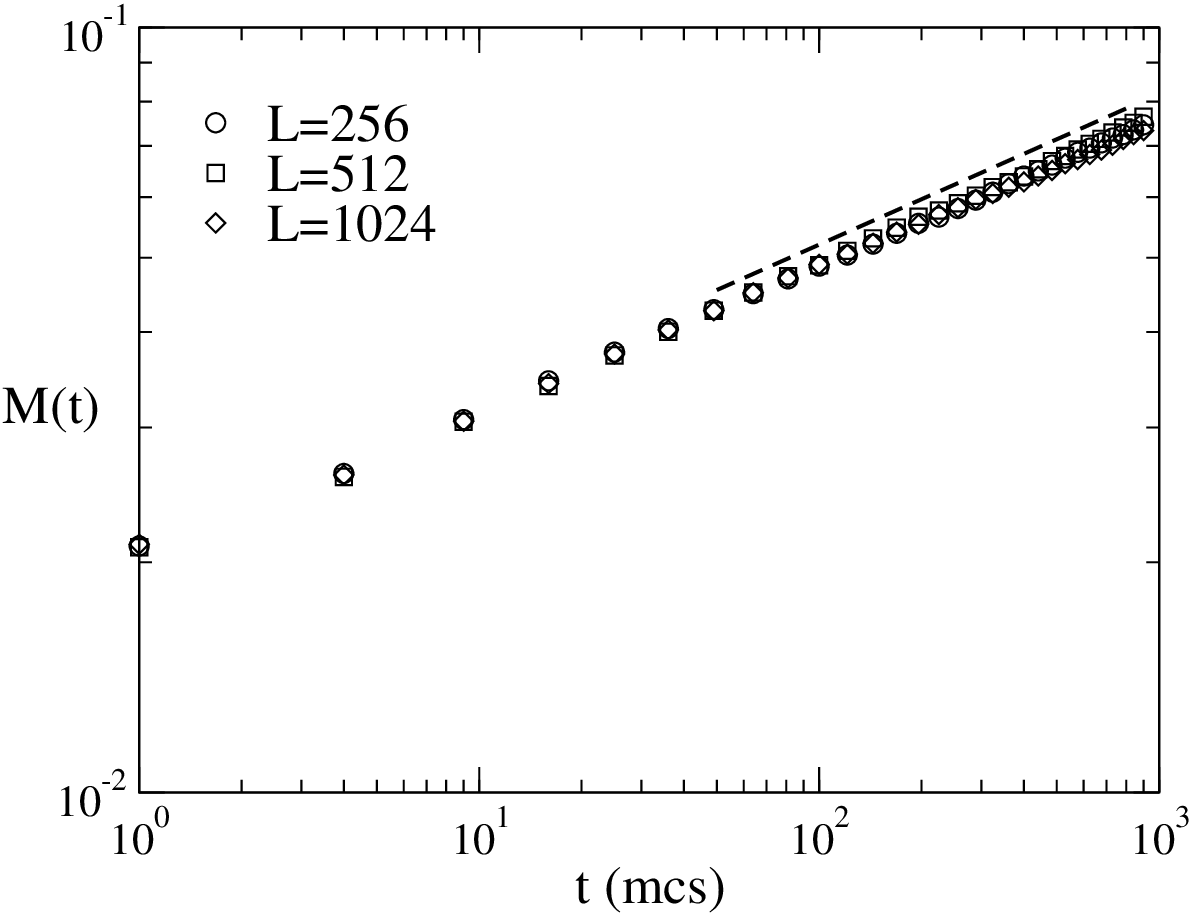}
\caption{\label{fig1} Time evolution of magnetization at the critical temperature $T_C$ for the Ising model by applying Metropolis algorithm. Results were obtained for $M_0=10^{-2}$ and different lattice sides $L$. The dashed line corresponds to the best fit of the data and has slope $\theta = 0.196(1)$.}
\end{figure}

In order to test the computational protocol, in Fig. \ref{fig1} we show the results obtained for the initial increase in the magnetization as a function of time (Eq. (\ref{eq:magnsh})), for lattices of different size $L$, $T=T_C$, and using Metropolis algorithm. In this example, $M_0=0.01$ is assumed. The best fit of the data yields $\theta=0.196(1)$ in excellent agreement with the best available value for the Ising model--with Metropolis dynamic--in the $M_0 \rightarrow 0$ limit, namely, $\theta=0.197(1)$ \cite{zheng98}.

\begin{figure}[!tbp]
\includegraphics[width=8cm,clip=true]{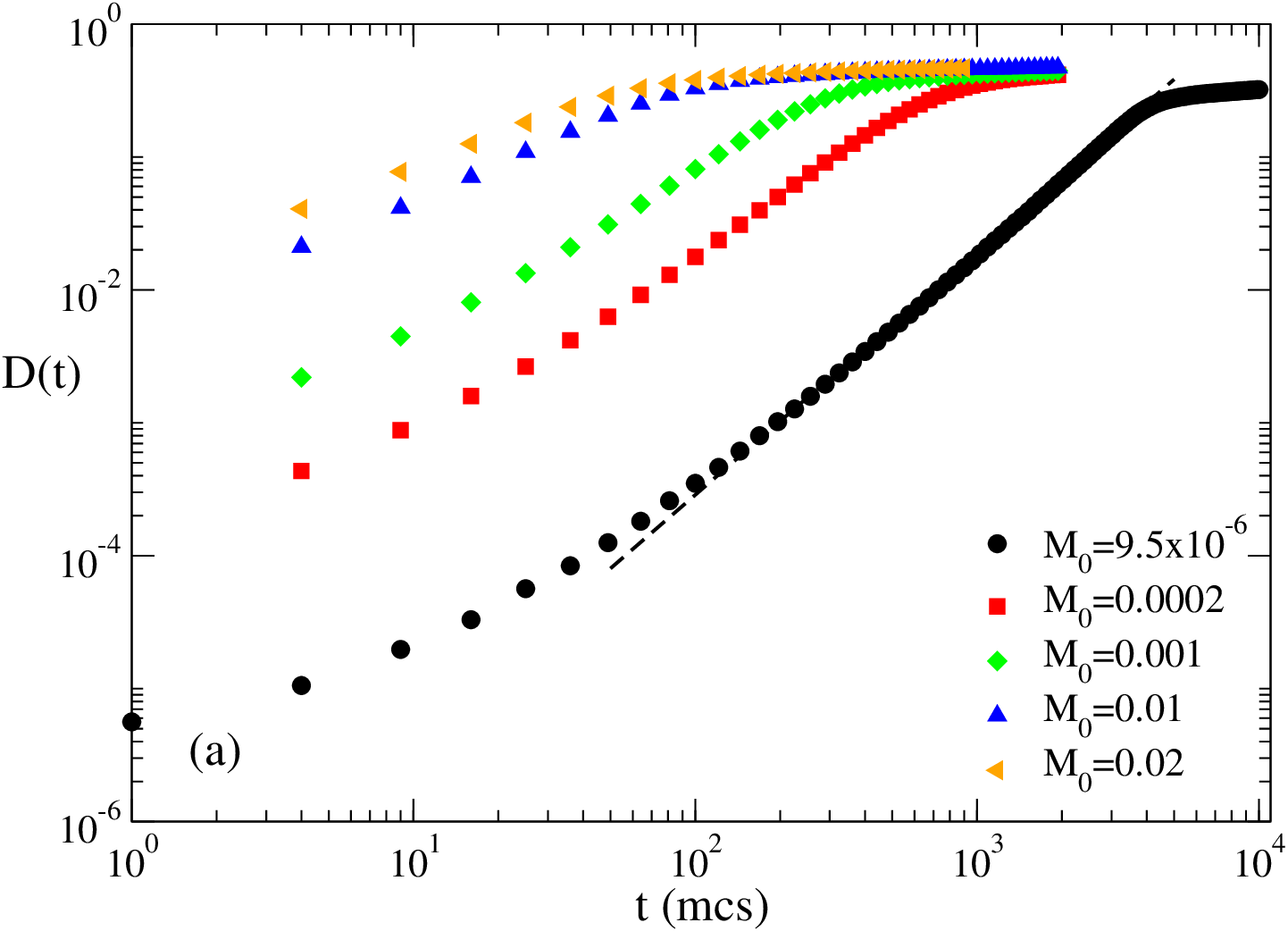}
\includegraphics[width=8cm,clip=true]{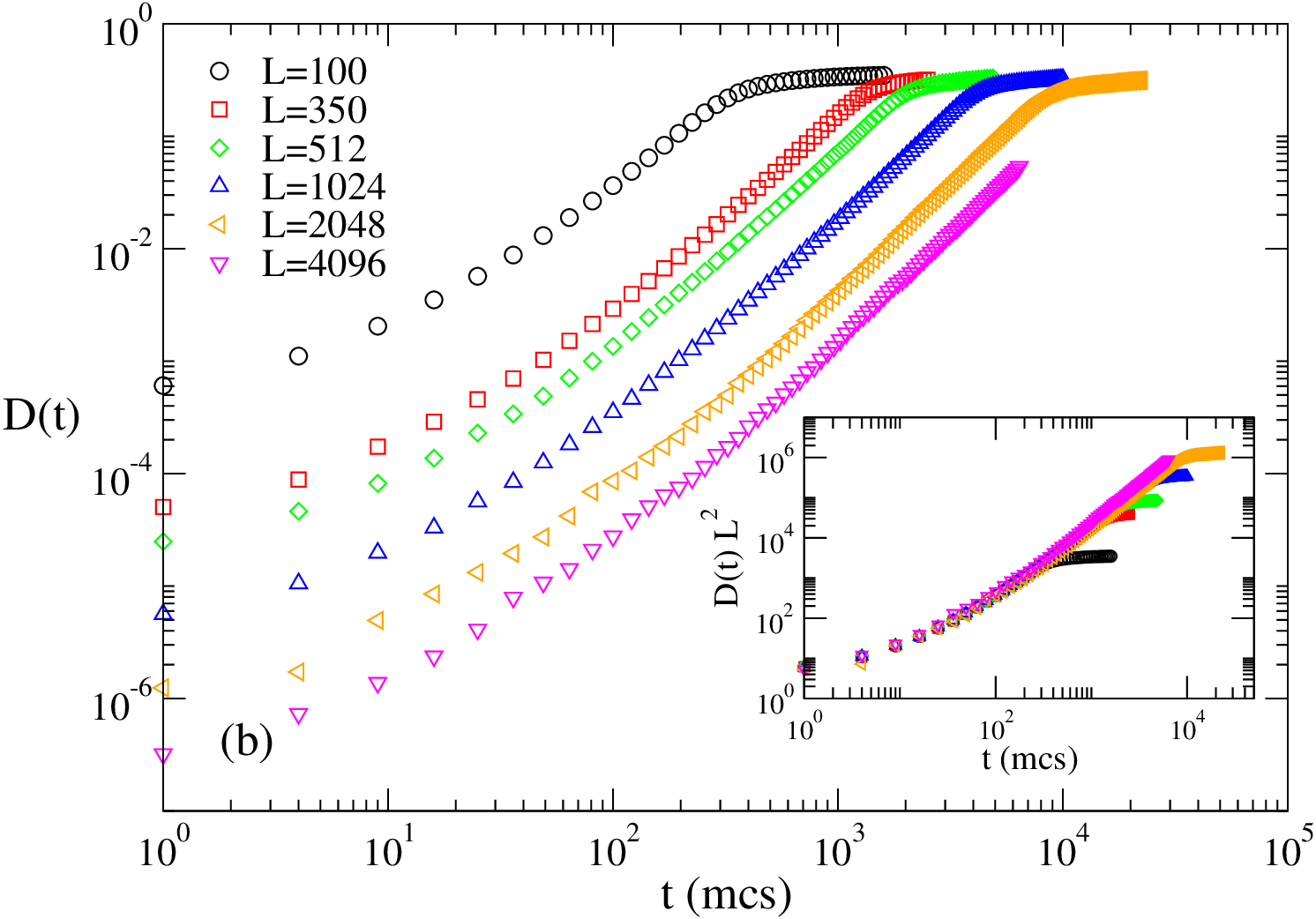}
\caption{\label{fig2} (Color online) (a) Log-log plot of damage vs $t$, obtained for a lattice of side $L=1024$, by applying the Metropolis algorithm and keeping $T=T_C$ constant. Different values of the initial magnetization $M_0= 2 D(0)$ are used, as listed in the main panel. The dashed line has slope $\theta_D(L)= 1.84(1)$.
(b) Damage vs $t$ for the epidemic study, with $M_0=2 \times 5/L^2$, $T=T_C$ and different lattice sizes, as shown in the figure. Inset: Collapse of the data.}
\end{figure}

\begin{figure}[!tbp]
\includegraphics[width=7cm,clip=true]{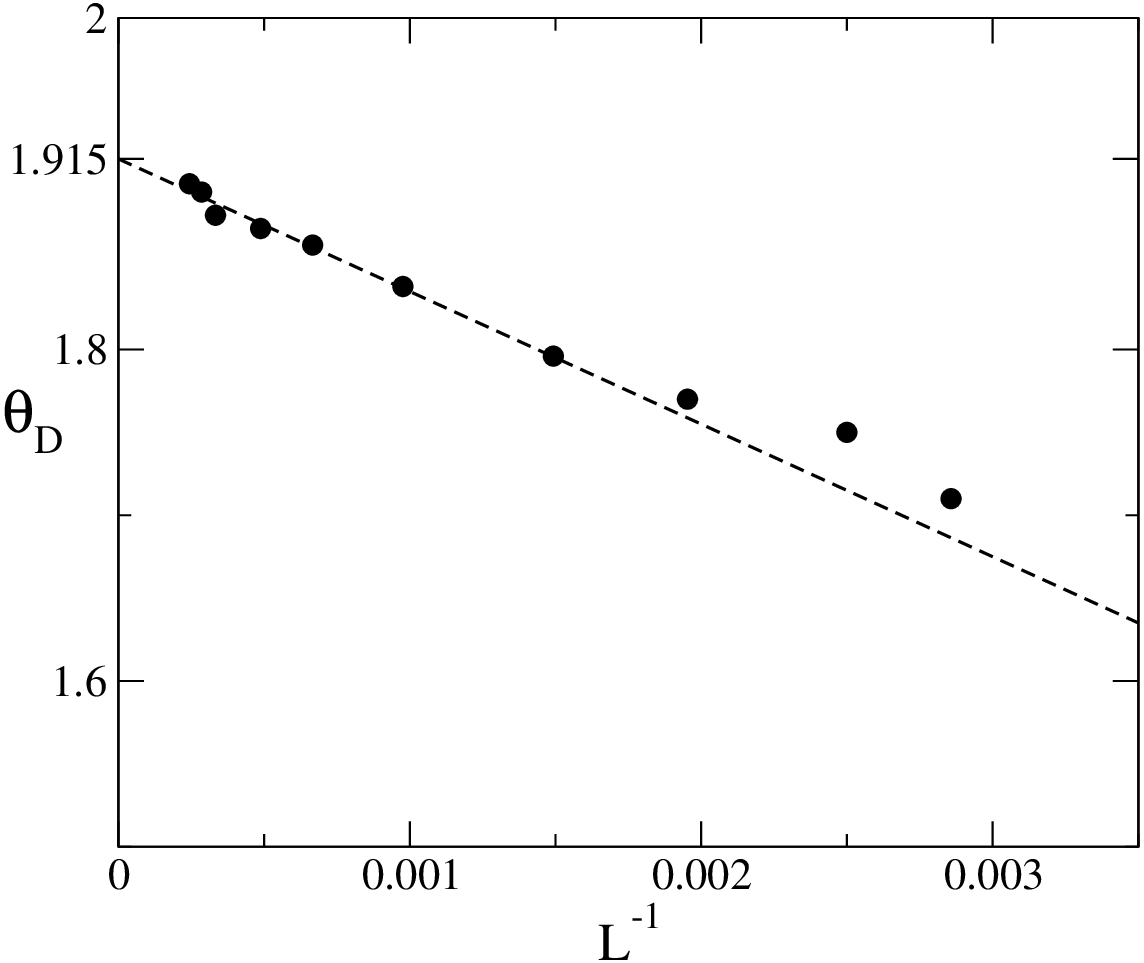}
\caption{\label{etaL} Linear behavior of the exponent $\theta_D$ as a function of $L^{-1}$, the inverse of the lattice size. The asymptotic limit values $\theta_D(L \rightarrow \infty) =1.915(3)$.}
\end{figure}

On the other hand, Fig. \ref{fig2}(a) shows plots of $D(t)$ versus $t$ obtained by keeping $T=T_C$ constant and varying the initial magnetization $M_0$ (and therefore the initial damage $D(0)$), for a lattice of size $L=1024$. The dashed line corresponds to a power-law behavior (Eq. \ref{eq:pwlaw}), but the results obtained show that the value of the exponent $\theta_D$ depends on the initial magnetization ($M_0$). 
The obtained values of $\theta_D$ also depend on the lattice size (not shown here for the sake of space), e.g., 
for a lattice side $L=1024$ and $T=T_C$, $\theta_D (M_0=10^{-2})=0.81(1)$, $\theta_D (M_0=10^{-3})=1.21(1)$ and $\theta_D (M_0=9 \times 10^{-6})=1.84(1)$ (see Figure \ref{fig2}(a)), and for $L=512$, $\theta_D (M_0=10^{-2})=0.76(1)$, $\theta_D (M_0=10^{-3})=1.15(1)$ and $\theta_D (M_0=9 \times 10^{-6})=1.77(1)$. 
For this reason and in order to establish the value of this exponent in the limits $M_0 \rightarrow 0$ and $L\rightarrow \infty$, we develop an epidemic study by flipping only the five central spins of the lattice. So, the initial magnetization for the lattice of size $L\times L$ is $M_0=2 D(t=0)=2 \times 5/L^2$.
The results obtained are shown in Fig. \ref{fig2} (b) for different lattice sizes as indicated in figure and at the critical temperature $T_C$. The inset of Fig. \ref{fig2}(b) shows the scaled results. 
After a transient period of $t \sim 10^2$ mcs, the damage exhibits a power-law behavior, but the value of the exponent $\theta_D$ also depends on the lattice size $L$, as is shown in Fig. \ref{etaL}. 
The data obtained suggest a size dependence of the form $\theta_D(L) = \theta_D(L \rightarrow \infty) + a L^{-1}$, and the best fit of the data gives $\theta_D(L \rightarrow \infty) =1.915(3)$ (see Fig. \ref{etaL}).
As can be seen this exponent is much bigger (roughly one order of magnitude) than the exponent for the initial increase of the magnetization, $\theta = 0.197$ \cite{zheng98}.

\begin{figure}[!tbp]
\includegraphics[width=8cm,clip=true]{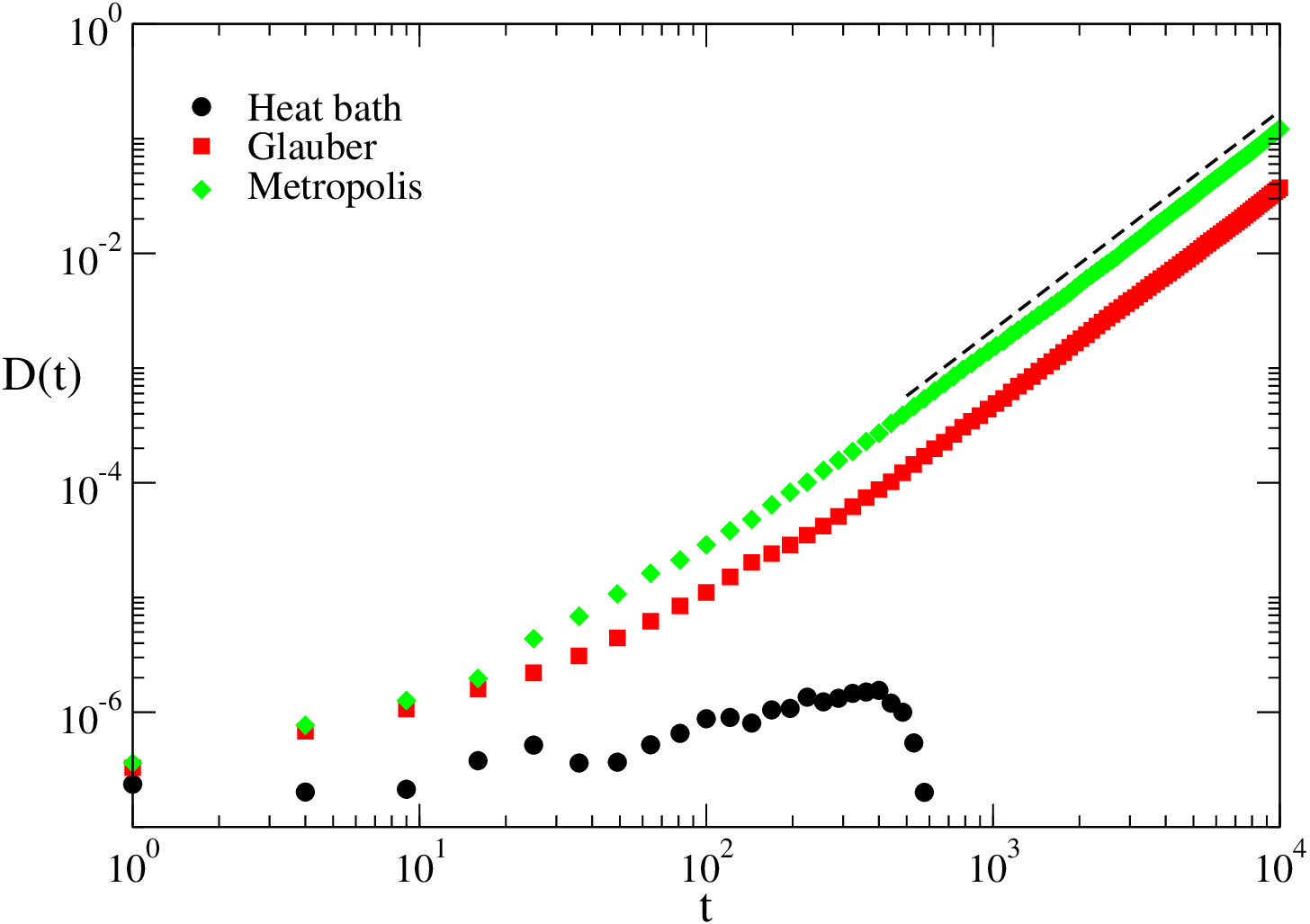}
\caption{\label{dynamic} (Color online) Log-log plot of damage as a function of time, for a lattice of size $L=4096$, $T=T_C$, $M_0=D(0)=5/L^2$, and different dynamic rules applied, as indicated in the figure.  The dashed line has slope $\theta _D \approx 1.91$.}
\end{figure}

In order to establish the dependence of this exponent on the dynamic rules applied to the lattice, we also performed simulations of the short-time dynamics of damage with heat bath, Glauber and Metropolis dynamics, in the epidemic case. The results obtained are shown in Fig. \ref{dynamic}. As is expected, the damage heals for heat bath dynamics at $T=T_C$, and it spreads in the cases of Metropolis and Glauber dynamics. In both cases, the value of the exponent is given by $\theta_D=1.91(1)$, in complete agreement with our previous result.

\begin{figure}[!tbp]
\includegraphics[width=8cm,clip=true]{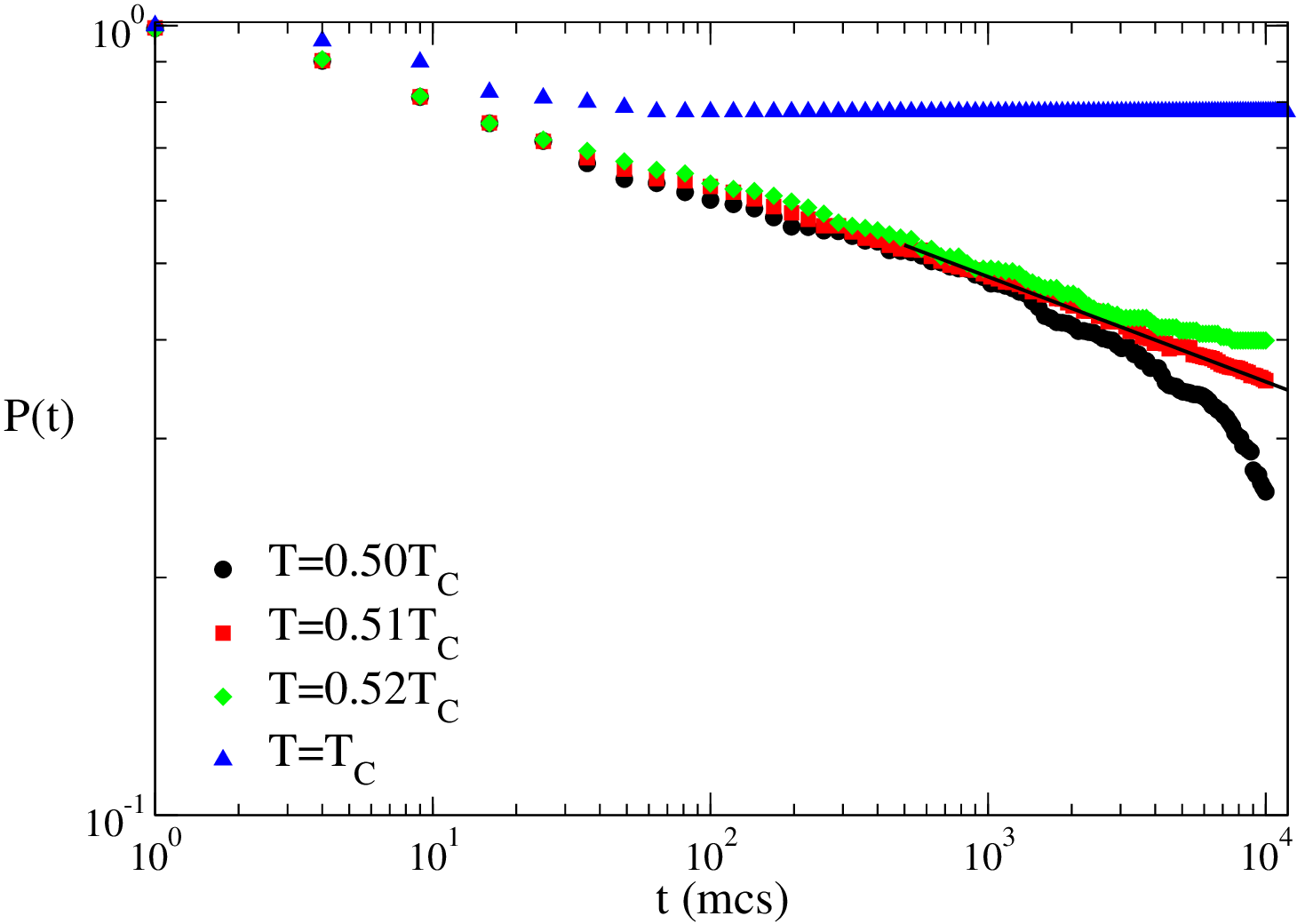}
\caption{\label{psurT} (Color online) Log-log plot of the survival probability of the damage ($P(t)$) vs $t$
obtained for a lattice size $L=2048$, and different temperatures $T$, as indicated in the figure. As can be observed, at $T=T_C$, $P(t)$ reaches a stationary value corresponding to $\delta=0$. The full line ($T=T_D\approx 0.51T_C$) has slope $\delta =0.133(1)$.}
\end{figure}

On the other hand, we analyze the dynamic dependence of the damage survival probability ($P(t)$).
Figure \ref{psurT} shows the results obtained for different temperatures and $L=2048$. 
As can be observed, $P(t)\rightarrow 0$ for $T\ll T_C$, while $P(t)$ reaches a stationary value at $T_C$. This behavior suggests that the damage transition occurs at a temperature lower than $T_C$, and at $T_C$ the decay exponent (see Eq. (\ref{probdelta})) is given by $\delta =0$. This result is in agreement with that reported by Montani et al. \cite{albmon} in the two-dimensional Ising model with Glauber dynamics.

On the other hand, we also studied the average distance from the origin of the propagation of damage, defined as \cite{grass}
\begin{equation}
\langle R^2(t)\rangle = \frac{\sum _{x}\langle (S^{A}_{x}(t,T)-S^{B}_{x}(t,T)) r^2\rangle}{\sum _{l}\langle (S^{A}_{x}(t,T)-S^{B}_{x}(t,T))\rangle},
\end{equation}
where $r=\left|(x-L/2)\right|$.

\begin{figure}[!tbp]
\includegraphics[width=8cm,clip=true]{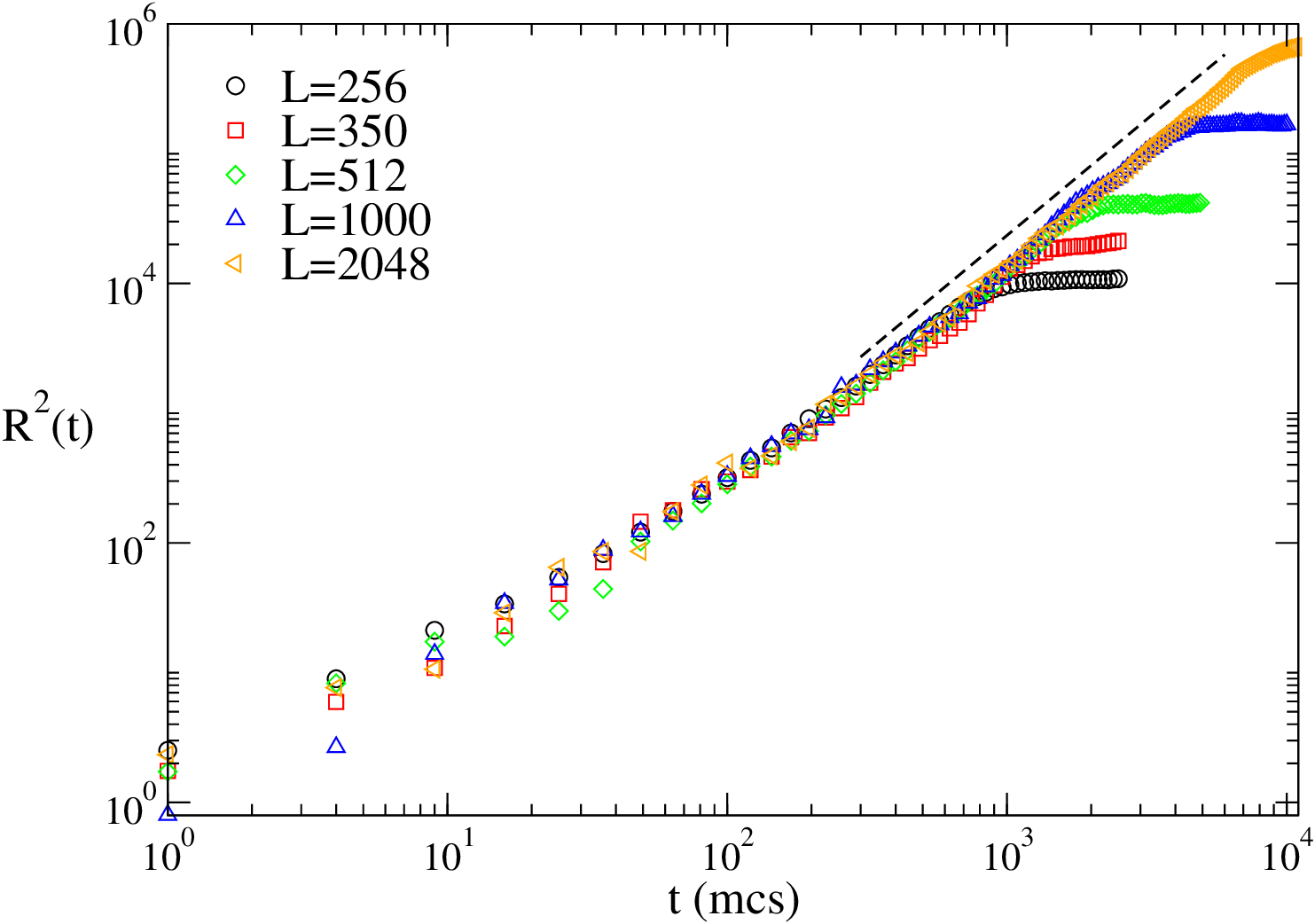}
\caption{\label{r2vst} (Color online) Average distance from the origin of the propagation of damage $R^2(t)$ vs $t$, for $T_C$ and different values of the lattice size $L$. The dashed line has slope $z=1.92$.}
\end{figure}

Figure \ref{r2vst} shows a log-log plot of $\langle R^2(t)\rangle$ versus $t$, for lattices of different size $L$ and $T=T_C$. These data suggest a power-law behavior for $\langle R^2(t)\rangle$ of the form
\begin{equation}
\langle R^2(t)\rangle  \approx t^{z^*},
\label{r2}
\end{equation}
where the best fit of the data (in the limit $L\rightarrow \infty$) gives $z^*=1.92(4)$. This exponent is similar, within error bars, to the exponent $\theta_D$. 

In order to understand the spatiotemporal evolution of both the magnetization and the damage, 
we also analyzed the profiles along the $x$ direction. 
For this purpose, we studied a strip geometry of size $L\times M$ with strictly zero magnetization. The initial damage was introduced by flipping only three spins of the central column ($i=M/2$) in the configuration labeled $S^{B}_{i,j}(t,T_C)$, so $M_0=2 D(0)=2\times 3/(L\times M)$. Thus, we define the damage and magnetization profiles as
\begin{equation}
D(i,t)=\frac{1}{2L}\sum ^{L}_{j=1}\left| S^{A}_{i,j}(t,T_C)-S^{B}_{i,j}(t,T_C)\right|,
\label{damprof}
\end{equation}

\begin{equation}
M(i,t)=\frac{1}{L}\sum ^{L}_{j=1} S^{A}_{i,j}(t,T_C),
\label{magprof}
\end{equation}
\noindent where $S^{A}_{i,j}(t,T)$ and $S^{B}_{i,j}(t,T)$ are the reference and damaged configurations at site $l$ of coordinates $\{i,j\}$, respectively.  
By assuming these definitions, the profiles $M(i,t)$ and $D(i,t)$ represents the average magnetization and damage 
of the $i$th ($i = 1,...,M$) row of the system, which runs parallel to the 
$x$ direction, respectively.

\begin{figure}[!tbp]
\includegraphics[width=8cm,clip=true]{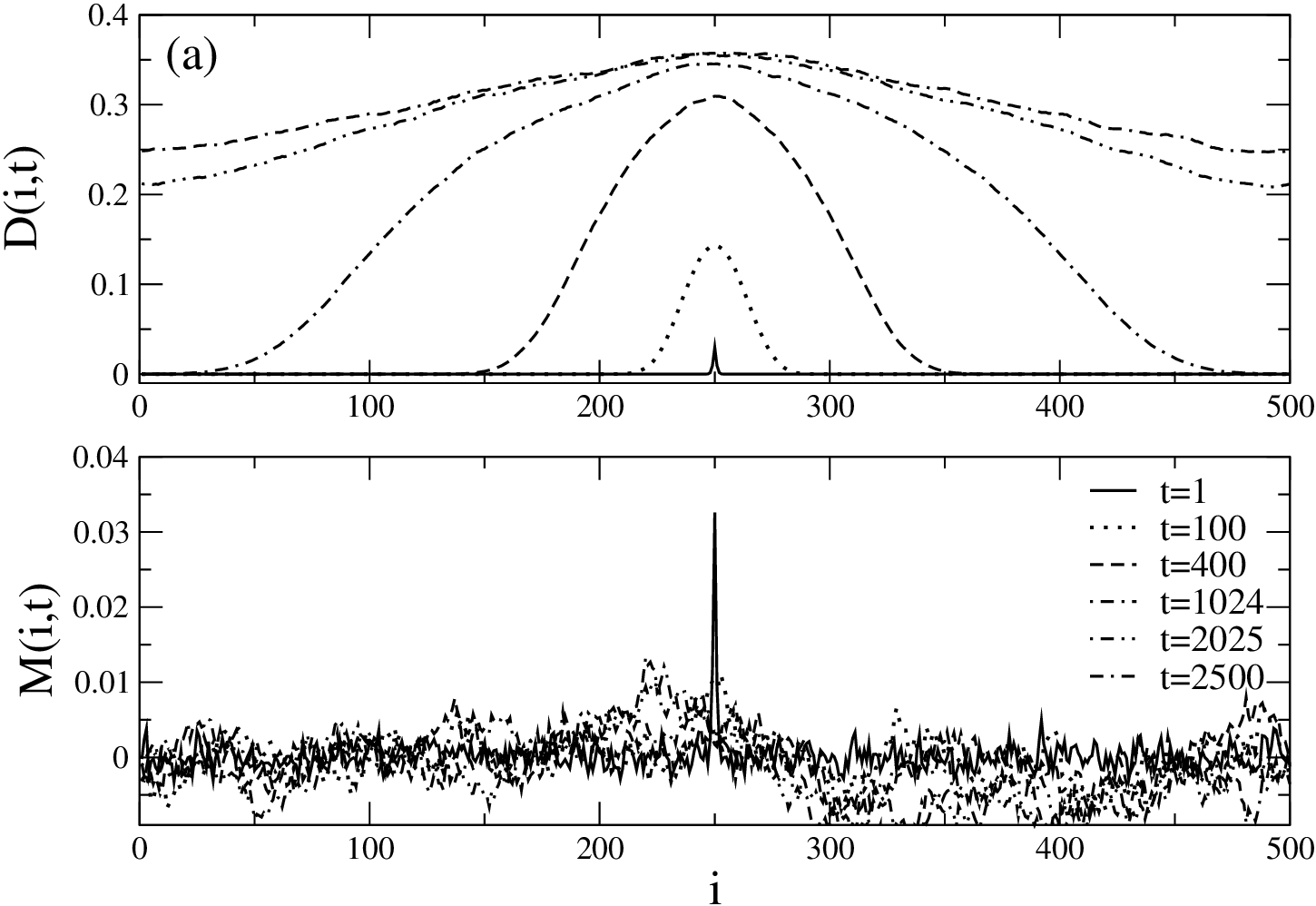}
\includegraphics[width=8cm,clip=true]{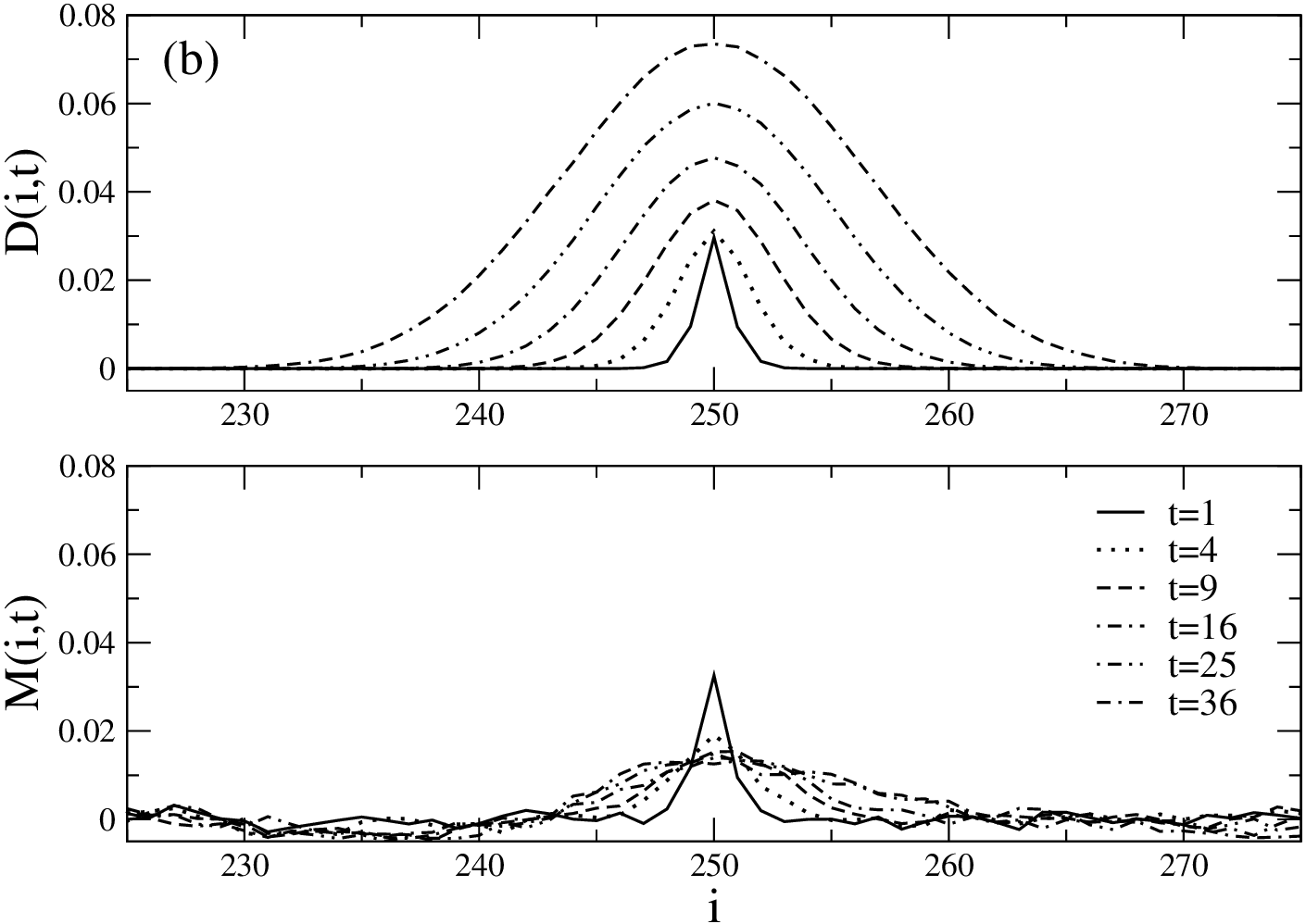}
\caption{\label{perfil} Damage and magnetization profiles at the $x$ direction for a lattice of size $L\times M=100 \times 500$. The results are obtained using Metropolis algorithm, at $T=T_C$ and for different times:
(a) $t=1, 100, 400, 1024, 2025$, and $2500$ mcs; (b) the initial times $t=1, 4, 9, 16, 25$, and $36$ mcs. 
The full line indicates, in both cases, the initial damage (or magnetization).}
\end{figure}

Figure \ref{perfil}(a) shows the results obtained for a lattice of size $L\times M=100\times 500$, at $T=T_C$ and different times as indicated in the figure caption. As can be observed, the propagation of damage is faster than that of the magnetization. This result is in complete agreement with the fact that for the initial increase exponents one has $\theta_D > \theta$. 
Also, in Fig. \ref{perfil}(b) we show, in more detail, the profiles for the initial times of the propagation, namely, $t<100$ mcs.

\begin{figure}[!tbp]
\includegraphics[width=8cm,clip=true]{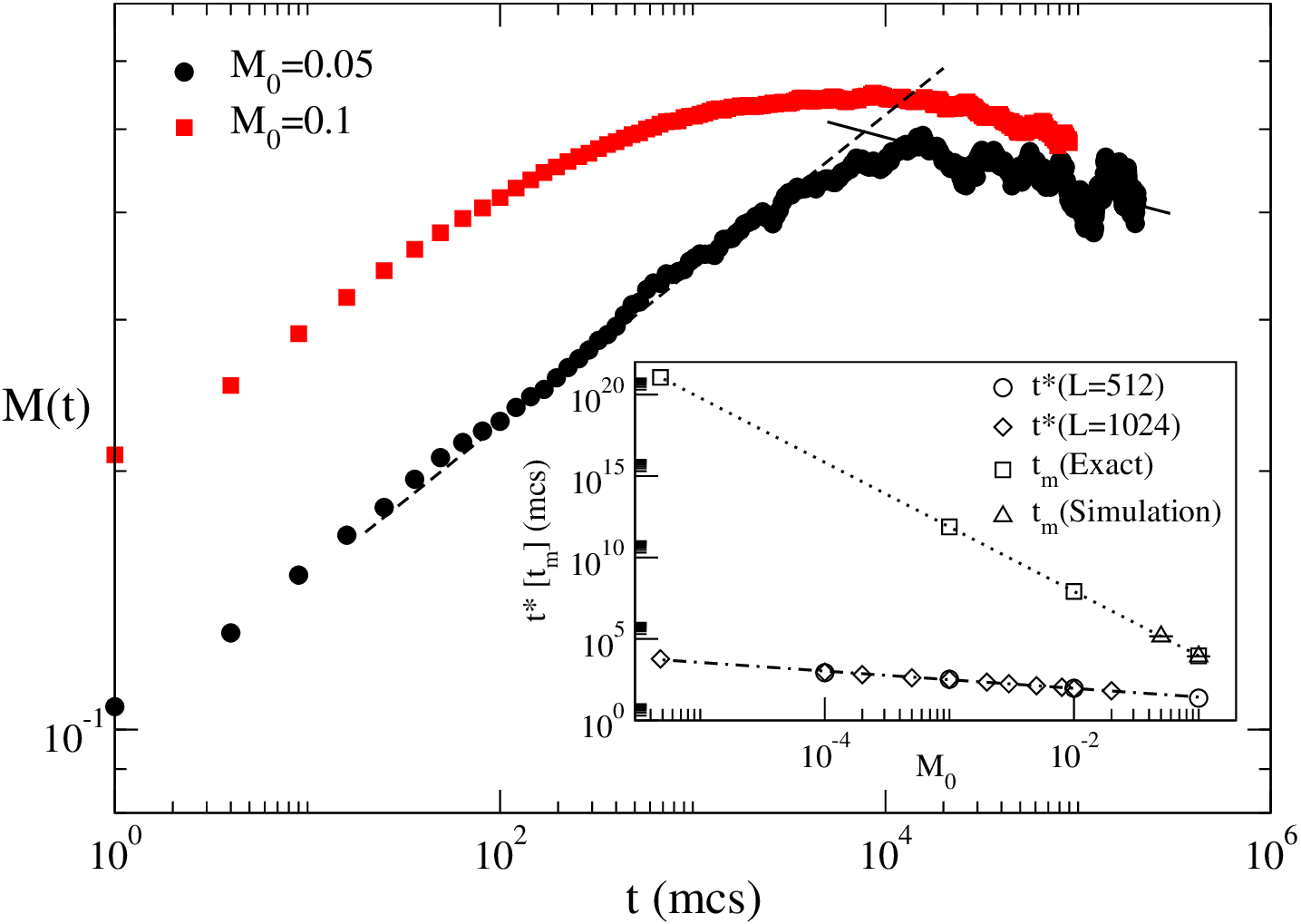}
\caption{\label{tmesos} (Color online) Time evolution of the magnetization at the critical temperature $T_C$ for the Ising model with $M_0=0.05$ and $0.1$. The dashed line corresponds to the initial increase $M \sim t^{\theta}$, with $\theta \approx 0.19$, and the full line shows the standard relaxation of $M(t)$ after a mesoscopic time $t_m$, namely $M \sim t^{-\beta/\nu z}$, with $\beta/\nu z = 0.0625$. The inset shows the crossover time of damage ($t^*$) and magnetization ($t_m$) vs $M_0$. The dotted line has slope $-z/x_0 \approx -3.96$, while the slope of the dash-dotted line is $-0.53(1)$.}
\end{figure}

As was mentioned in Sec. \ref{intro}, after a mesoscopic time $t_m \sim M_0^{-y}$ ($y=z/x_0$), the magnetization decays as $M \sim t^{-\beta/\nu z}$ (see Fig. \ref{tmesos}). For the Ising model $y \approx 3.96$ and $\beta/\nu z \approx 0.0625$, so that for an initial magnetization $M_0=0.1$, $t_m \sim 10^4$mcs, $M_0=0.05$, $t_m \sim 10^5$mcs, and $M_0=10^{-5}$, $t_m \sim 10^{21}$mcs.
On the other hand, after a certain time $t^*$, the damage reaches a saturation value. This characteristic time $t^*$ was determined as a function of lattice size and the initial magnetization $M_0$ (see the inset of Fig. \ref{tmesos}).
As can be seen, there are no finite size effects (at least for the lattice sizes used in the simulations). 
Moreover, the inset of Fig. \ref{tmesos} shows a power-law behavior of the form $t^*\sim M_0^{-y_D}$, with $y_D=0.53(1)$. 
By comparing this value with the magnetization given by $y \approx 3.96$ (squares in the inset of Fig. \ref{tmesos}), it can be observed that the damage spreads over the lattice much more quickly than the magnetization, in agreement with previous results shown in this paper. 

\begin{figure}[!tbp]
\includegraphics[width=8cm,clip=true]{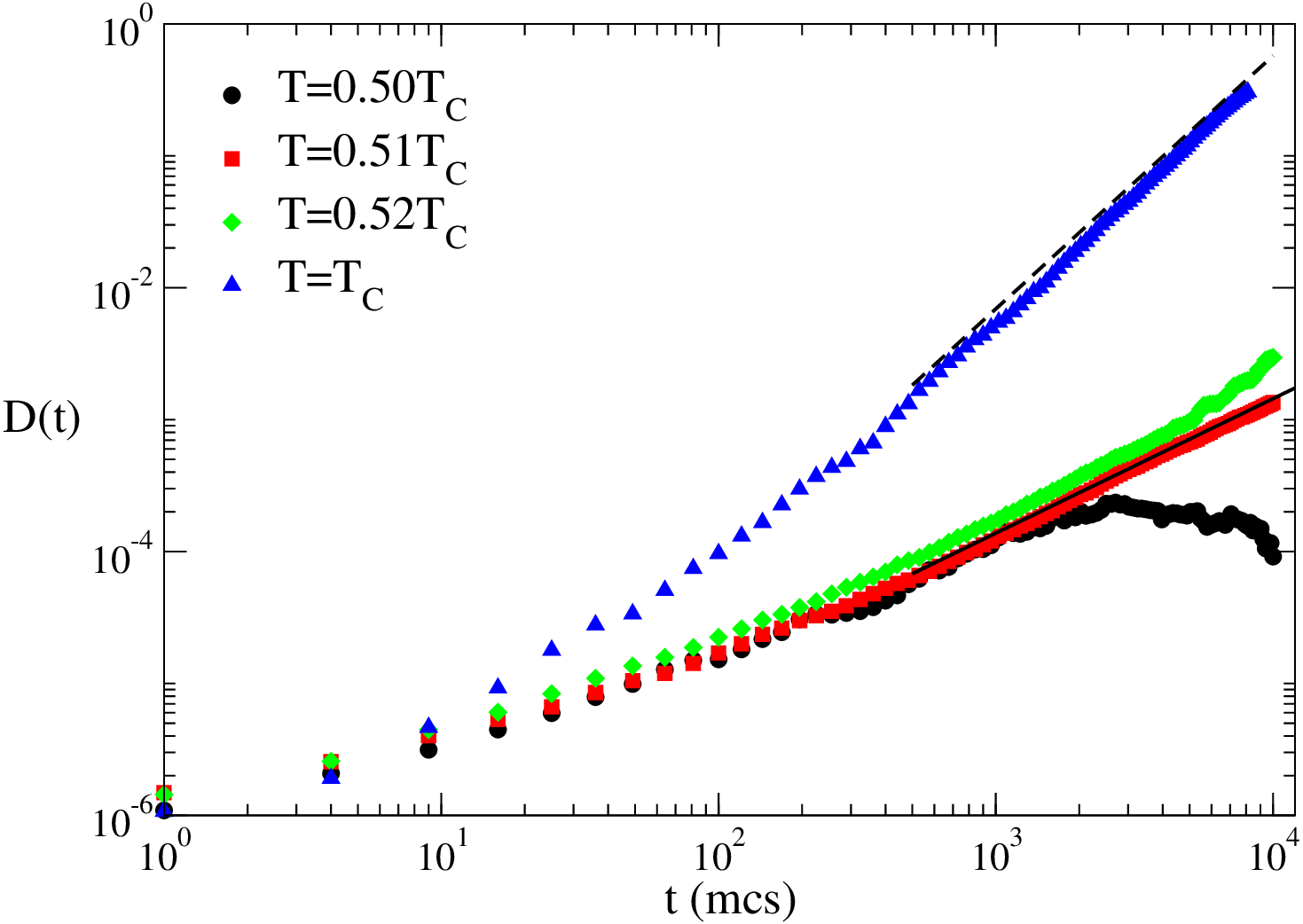}
\caption{\label{damT} (Color online) Log-log plot of the damage $D(t)$ vs $t$ obtained for a lattice size $L=2048$ and different temperatures $T$ as indicated in the figure. The dashed line (corresponding to $T=T_C$) has slope $\theta_D(L=2048)= 1.91$, and the full line ($T=T_D\approx 0.51T_C$) has slope $\theta_D =1.026(3)$.}
\end{figure}

Finally, the behavior observed for the survival probability (see Fig. \ref{psurT}) suggests that the critical temperature for damage transition is lower than $T_C$. 
In fact, there is another (lower) temperature at which both $D(t)$ and $P(t)$
behave as power laws. 
This happens at $T_D=0.51(1)T_C$, and the corresponding exponents are
$\theta_D = 1.026(3)$ and $\delta = 0.133(1)$ (see Figs. \ref{psurT} and \ref{damT}). 
Also, the exponent $z^*=1.74(3)$ was determined at $T_D$.

\section{Conclusions}
\label{sec4}

The propagation of damage and the short-time critical dynamics of the $d=2$ Ising model are studied.
For this purpose, fully uncorrelated initial configurations ($T=\infty$) are suddenly quenched at the measured temperature.
The proposed methodology for the study of the spreading of the damage is in contrast to previous studies \cite{rubio2008} where the equilibration and measurement temperatures are identical.
It is found that by quenching the system at $T_C$, or close to it, the spreading of the damage is much faster than the initial increase of the magnetization.
However, the critical temperature for damage spreading is far below $T_C$, namely close to $T_D=0.51(1)T_C$.
It is worth mentioning that the values of the exponents measured at $T_D$, namely, $\theta_D = 1.026(3)$, $\delta = 0.133(1)$, and $z^*=1.74(3)$, are different from those reported by Grassberger for the Ising model with heat bath dynamics at $T_C$ \cite{grass}.
However, it must be mentioned that both systems are quite different. 
In our case, the damage was introduced in a completely uncorrelated configuration that was suddenly quenched at a temperature $T$. 
Subsequently, the damage starts to evolve with the standard Metropolis dynamics at $T$.
Therefore,  a completely different scenario is expected: the critical behavior is not observed at $T_C$--or close to it, as usual--but at a much lower temperature given by $T \approx 0.51 T_C$. 
This result would mean that the damage spreading Transition observed in our case is completely different, and for this reason, the values of the critical exponents reported here are not the same as those reported by Grassberger \cite{grass}.

Finally, we would like to remark that our results may indicate that the damage spreading critical temperature would systematically decrease when increasing the equilibration temperature, $T_D=0.51(1)T_C$ being the lower bound as obtained for $T_{equil}=\infty$.

\section*{Acknowledgments}
This work was supported financially by CONICET, UNLP, and ANPCyT (Argentina). 

\bibliographystyle{unsrt}
\bibliography{damST}

\end{document}